\documentclass[fleqn,twoside,10pt]{article}
\usepackage{espcrc2,amssymb,epsfig,amsmath,amsfonts}

\usepackage[figuresright]{rotating}

\newcommand{\Tr}{\text{\,Tr\,}}
\newcommand{\LL}{\mathcal{L}}

\newcommand{\nlsm}{nl$\sigma$m }

\title{Little Higgs Models: New Approaches to the Hierarchy Problem}

\author{J. G. Wacker\address[Harvard]{Theory Group, Department of Physics\\University of California, \\ 
        Berkeley, CA 94720 USA}%
        \thanks{Theory Group, Harvard University}
}
\begin{document}

\begin{abstract}
In this note we present a review of the little Higgs models that 
stabilize the electroweak by realizing the Standard
Model Higgs as a pseudo-Goldstone boson.
\vspace{1pc}
\end{abstract}

\maketitle

\section{Introduction}

The experimental results of the past decade indicate that the
chiral Lagrangian describing EWSB works to one loop accuracy
and irrelevant operators (such as $|D \omega|^4$, where $\omega$
are the Goldstone bosons eaten by $W^\pm$ and $Z^0$) are suppressed
by a scale $\Lambda \gg 1 \mbox{ TeV}$.  This strongly indicates perturbative
physics at $1 \mbox{ TeV}$ and a physical Higgs boson $H = ( h^0 + v) \omega$.
However, this linear sigma model is unstable to quantum corrections leading
to the hierarchy problem, meaning that the Standard Model is in an
incomplete description of physics parametrically above the weak 
scale.  To date, only the MSSM provides a solution to the hierarchy problem
and weakly coupled physics at 1 TeV.

In this note we review models where the Higgs is a pseudo-Goldstone
boson, $\Sigma= e^{i \sigma/f}$ \cite{Arkani-Hamed:2001nc}, for early
attempts see \cite{Kaplan:1983fs}.  
Approximate global symmetries, $\Sigma \rightarrow L \Sigma R^\dagger$
keep the Higgs vev small relative to the cut-off, $\Lambda \sim 4 \pi f$.  
The global symmetries are  broken by couplings, however,
the structure of the symmetry breaking prevents large
corrections to the Higgs vev.  At 100 GeV the model is a two Higgs
doublet model with a Higgs potential identical to the MSSM.  
There is a mini-desert up to 1-3 TeV where new
particles emerge that cancel the quadratic divergences of the 
Standard Model.  The novelty of this set of models is that particles
of the \begin{it}same\end{it} spin cancel the respective quadratic divergences.
Above the TeV scale there is another mini-desert until the 10-30 TeV scale
where the description of the model becomes strongly coupled and new
physics emerges.  However, the Standard Model is insensitive to this
physics.

\vspace{-0.4cm}
\begin{figure}[ht]
\centering\epsfig{figure=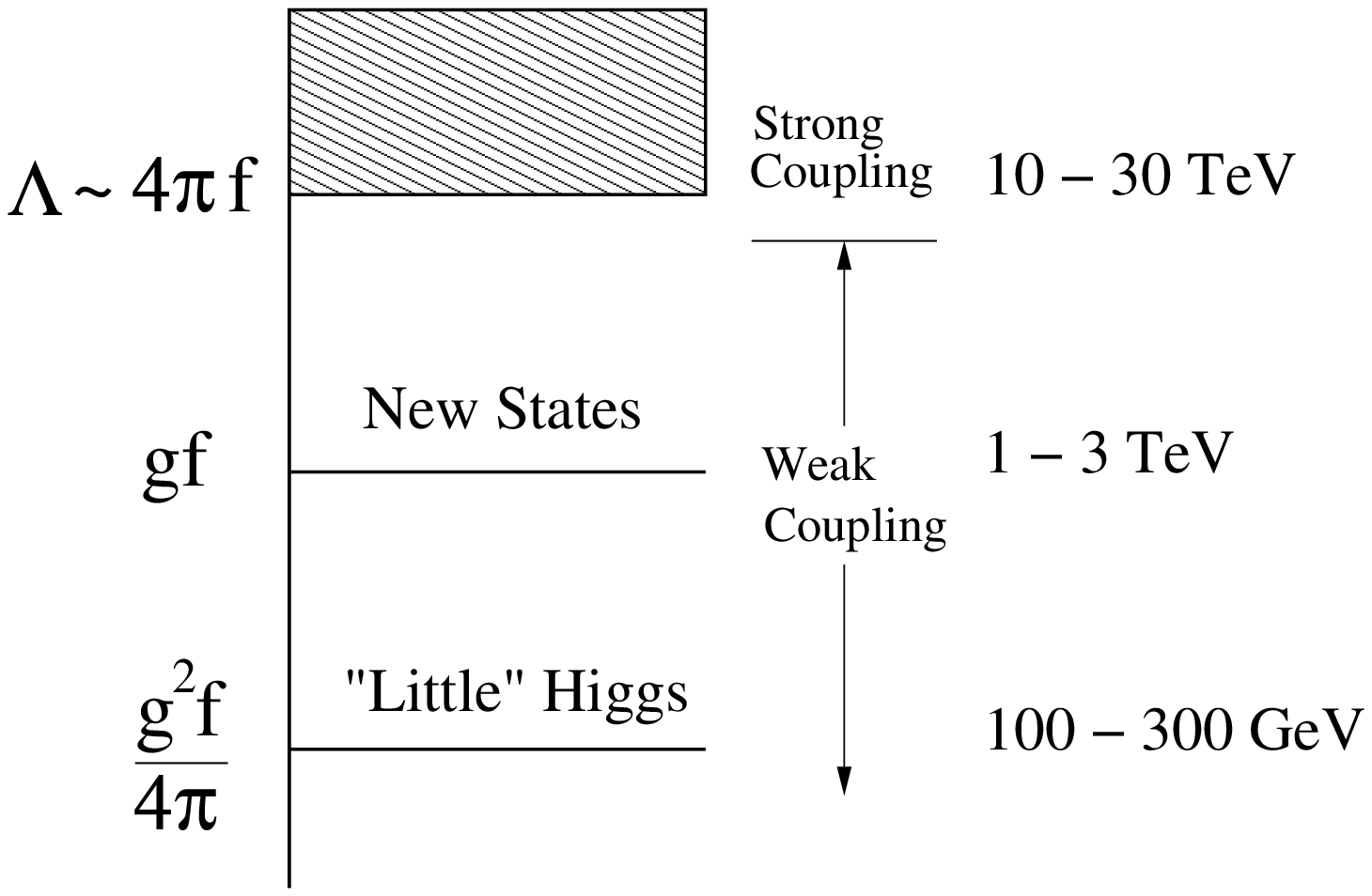, width = 7cm}
\end{figure}
\vspace{-0.7cm}

\section{Little Higgs Models}

Little Higgs models are gauged non-linear sigma model with
Lagrangians that can be divided into four pieces:
\begin{equation}
\hspace{-0.2cm}\LL = \LL_{\text{Gauge}} + \LL_{\text{nl$\sigma$m Kin}} 
+ \LL_{\text{Plaq}} + \LL_{\text{Yuk}}
\end{equation}

The first term is the standard kinetic term for gauge bosons.
The second term is the \nlsm kinetic terms that give mass to 
some of the gauge bosons.  There is the possibility of adding
a scalar potential, often called the plaquette potential, that gives mass to 
some of the \nlsm fields.  
Finally there are the Yukawa couplings between the \nlsm fields and fermions.
To study the low energy dynamics of the
theory we want to integrate out all the states that are classically
massive and have an effective Lagrangian in terms of the classically
massless fields.  
Some \nlsm can be represented by theory spaces where points denote
gauge groups, lines are \nlsm fields, and faces are plaquette potentials.
This was studied in depth in \cite{Gregoire:2002ra}.
The classically massless scalars are 
the little Higgs and are associated with non-contractible loops
in theory space.  The little Higgs has a mass generated
radiatively, but the size of the radiative corrections are parametrically
smaller than the TeV scale. 


In the limit of vanishing couplings the \nlsm fields become exact Goldstone
bosons under the symmetry $\Sigma \rightarrow L \Sigma R^\dagger$.  This
leaves the little Higgs exactly massless, but also without non-derivative
interactions.  Coupling constants break these chiral symmetries and
allow masses for the little Higgs to be generated radiatively.
Little Higgs models always require at least two couplings to communicate
sufficient chiral symmetry breaking to generate a mass radiatively,
and therefore quadratic divergences first arise at two loop order.
To ensure that there are no one loop quadratic divergences
to the the little Higgs mass there are simple rules for theory space
models \cite{Arkani-Hamed:2002qx,Gregoire:2002ra}.  
The first rule is that if all \nlsm fields are bi-fundamentals, then
there are no one loop gauge quadratic divergences.
If no plaquette contains any \nlsm field more than once, then there 
one loop quadratic divergences to the little Higgs mass from the scalar 
potential.

Having removed the one loop quadratic divergences, there are always 
finite contributions to the little Higgs mass.  We can estimate
the size of these finite contributions by studying the theory
beneath the scale of the classically massive modes.
In this low energy thoery there are a quadratic divergences and 
they are cut-off by the mass of the first heavy states and become 
finite corrections to the mass squared of the little Higgs.  
The mass of the heavy modes is roughly
$M_{\text{H}} \sim g f$.  This means that the correction to the little
Higgs mass from the low energy quadratic divergence is:
\begin{eqnarray}
\delta m^2_{\text{LH}} \sim \frac{g^2}{16 \pi^2} M^2_{\text{H}} 
= \frac{g^2}{16 \pi^2} g^2 f^2 
= \left( \frac{g^2}{16 \pi^2} \right)^2 \Lambda ^2 .
\end{eqnarray}
This is the same size as the two loop quadratic divergence in the
full theory.  Thus, even if we remove higher order quadratic divergences,
we can not push the cut-off higher than two
loop factors above the EW scale, setting  $\Lambda \sim 10 - 30 \text{ TeV.}$  
This still allows for a seperation of scales
\mbox{$m_{\text{LH}} \ll M_{\text{H}} \ll \Lambda.$}
Furthermore, this means that the states that stabilizes the electroweak
scale is invisible to roughly $1 \text{ TeV.}$

\section{The Minimal Moose}

\begin{figure}[ht]
\centering\epsfig{figure=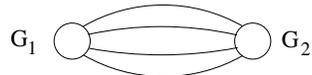, width = 4.0cm}
\vspace{-0.7cm}
\caption{\label{Fig: Minimal} Minimal theory space model.}
\end{figure}

The minimal theory space model\footnote{
see \cite{Arkani-Hamed:2002qy,Low:2002ws} for other classes of models} 
that has the requisite properties for EWSB has two gauge group and 
four \nlsm fields and is illustrated in Fig. \ref{Fig: Minimal}.   
The two gauge groups are taken to be $SU(3)$ and $SU(2)\times U(1)$ with
the $U(1)$ embedded as the $\mathbf{T_8}$ generator of 
$SU(3)$.  There are two plaquettes added:
\begin{eqnarray}
\nonumber
\hspace{-0.3cm}
\LL_{\text{Plaq.}} &=&
\lambda_1 f^4 \Tr X_1 X_2^\dagger X_3 X_4 ^\dagger + \text{ h.c.}\\
&&+ \lambda_2 f^4 \Tr X_2 X_3^\dagger X_4 X_1 ^\dagger +\text{ h.c.  } 
\end{eqnarray} 
This gauged \nlsm breaks the $SU(3)\times [ SU(2)\times U(1)]$
gauge symmetry down to the diagonal $[SU(2)\times U(1)]$ that is identified
as the electroweak gauge group.    
The plaquettes give mass to one linear combination of \nlsm fields and
leave three massless.  One is eaten by the massive $SU(3)$ gauge bosons
while two are physical and massless and are the little Higgs of the theory.

\vspace{-0.4cm}
\vspace{-0.4cm}

Since all the link fields are bi-fundamentals under the two gauge groups
and the plaquettes contain each link field only once, there are no one
loop quadratic divergences to the little Higgs masses.

The scalar potential for little Higgs, $\Sigma_{1,2}$, is of the
form $ \Tr \Sigma_1 \Sigma_2 \Sigma_1^\dagger \Sigma_2^\dagger$ or 
in terms of the linearized modes:
\begin{equation}
V(\sigma_1,\sigma_2) = \lambda_{\text{eff}} \Tr [ \sigma_1, \sigma_2 ]^2 
+ \cdots
\end{equation}
where $\lambda_{\text{eff}}^{-1} = \lambda_1^{-1} + \lambda_2^{-1}$.
Written in terms of the doublets, the potential is:
\begin{eqnarray}
V(h_1, h_2) = \lambda_{\text{eff}} \left(|h_1|^2 - |h_2|^2\right)^2
\end{eqnarray} 
This is identical to the Higgs potential inside the MSSM except that the 
coefficient is undetermined.  Hence, the mass of the Higgs is undetermined,
although we do expect $\lambda_{\text{eff}} \sim \mathcal{O}(1)$, leading
to a physical Higgs mass in the \mbox{$100 - 300$ GeV} range. 

The Standard Model fermions are charged under the $SU(2)\times U(1)$ gauge
group.  It is convenient to write the fermions as triplets under the
global $SU(3)$, i.e. $Q = (q, 0)$ and $U^c = (0,0,u^c)$.
They are coupled to the little Higgs, 
$\Sigma = \exp(i(\sigma_1\pm \sigma_2))$, through operators such as:
\begin{eqnarray}
yf Q \Sigma U^c
\sim y q ( h_1 \pm i h_2) u^c + \cdots .
\end{eqnarray}
This operator induces a one loop quadratically divergent contribution
to the little Higgs mass.  For everything but the top quark, the quadratic
divergence is negligeable.   For the top quark, the one loop quadratic 
divergence must be dealt with.   By adding an additional massive vector-like
fermion, $\chi$ and $\chi^c$, we can promote the quark doublet, $q$, to
quark triplet under the global $SU(3)$, $Q = (q, \chi)$.  By preserving
the global symmetry of the coupling, we remove the one loop quadratic 
divergence to the top quark.  The quadratically divergent part of
the one loop Coleman-Weinberg effective potential is:
\begin{equation}
V_{\text{eff}} = - \frac{y^2\Lambda^2 f^2}{16 \pi^2} \Tr P_R \Sigma P_L \Sigma^\dagger
\end{equation}
where $P_R = (0,0,1)$ from the $U^c$ coupling and $P_L = (1,1,0)$ if 
without the vector-like fermion $\chi$ and becomes the identity matrix with
with $\chi$.   When $P_L$ is the identity matrix, we preserve the $SU(3)$
symmetry and the effective potential is independent of $\Sigma$.

One loop log divergences from the top sector drives one linear combination
of the Higgs to acquire a mass and the orthogonal combination are stabilized
by one loop log divergences from the gauge and scalar sectors 
\cite{Arkani-Hamed:2002pa,Arkani-Hamed:2002qy}.  
This leads to phenomenologically acceptable EWSB driven radiatively much 
as in the MSSM.  We refer the reader to \cite{Arkani-Hamed:2002qy} for
more details.

\section{Conclusions and Outlook}

We have presented new models of EWSB that stabilize the weak scale
to radiative corrections and have weakly coupled physics at 1 TeV.
At the 100 GeV scale it is a 2HDM model 
with possibly additional scalar singlets and $SU(2)$ triplets. 
At 1 TeV there are additional vector bosons that cancel the SMs 
guage $\Lambda^2$ divergences, additional scalars that cancel the
Higgs quartic interaction $\Lambda^2$ divergence and additional
coloured fermions that cancel the top's $\Lambda^2$ divergence.
This description is valid up to $10 - 30$ TeV where the \nlsm
becomes strongly coupled and new physics emerges.

There is a tremendous amount of physics that is still to be done:
precision studies \cite{Chivukula:2002ww}, experimental
signature \cite{Arkani-Hamed:2002pa}, exploring other little Higgs
models \cite{Arkani-Hamed:2002qy,Low:2002ws}, and UV completing
the theory above $10 - 30$ TeV.

\end{document}